# Eigensolutions of the N-dimensional Schrödinger equation interacting with Varshni-Hulthén potential model

E. P. Inyang, E. S. William and J. A. Obu

*Theoretical Physics Group, Department of Physics, University of Calabar, PMB 1115, Calabar, Nigeria*

*Corresponding author email: williameddyphysics@gmail.com*

Analytical solutions of the N-dimensional Schrödinger equation for the newly proposed Varshni-Hulthén potential are obtained within the framework of Nikiforov-Uvarov method by using Greene-Aldrich approximation scheme to the centrifugal barrier. The numerical energy eigenvalues and the corresponding normalized eigenfunctions are obtained in terms of Jacobi polynomials. Special cases of the potential are equally studied and their numerical energy eigenvalues are in agreement with those obtained previously with other methods. However, the behavior of the energy for the ground state and several excited states is illustrated graphically.

1. Introduction

The interactions of quantum systems with the spherically symmetric potentials have been studied by researchers since the discovery of quantum mechanics [1-3]. From the early days of quantum mechanics, the study of exactly solvable problems has attracted a considerable attention in many areas of physics, particularly in atomic physics, information theory, nuclear physics, particle physics, molecular physics and its important cannot be overemphasize [4,5]. The exact solution of relativistic and non-relativistic equation for most of this physical potential creates a problem where the application of the approximation is indispensable [6]. For instance, in the case of the Schrödinger equation, when the angular momentum quantum number is present, one can resort to solve the non-relativistic equation approximately via a suitable approximation scheme [7]. Some of such approximations, yielding good results, consist of the conventional approximation scheme proposed by Greene and Aldrich [8], the improved approximation scheme by Jia *et al.* [9], the elegant approximation scheme [10], the Pekeris approximation [11], the improved approximation scheme by Yazarloo *et al.* [12], improved approximation scheme in Refs. [13-15] and in Refs. [16, 17].

Over the past decades, problems involving the multidimensional Schrödinger equation have been addressed by many researchers using different analytical procedures. For examples, Ntibi *et al* [18] investigated the analytical solution of the D-Dimensional radial Schrödinger equation with Yukawa potential. Oyewumi *et al* [19] studied the N-dimensional Pseudoharmonic oscillator. Gönül and Koçak [20] investigated explicit solutions for N-dimensional Schrödinger equations with position-dependent mass. The N-dimensional Kratzer–Fues potential was discussed by Oyewumi [21], Ikhdair and Sever [22], studied the modified Kratzer–Fues potential plus the ring shape potential in D-dimensions while Dong [23], reviewed the wave equations in higher dimensions.

The Varshni potential is greatly important with applications, cutting across nuclear physics, particle physics and molecular physics. The Varshni potential model takes the form [24, 25]:

$$V_V(r) = \eta_0 - \frac{\eta_0 \eta_1}{r} e^{-\delta r}, \qquad (1)$$

where $\eta_0$ and $\eta_1$ are potential strength parameters, $\delta$ is the screening parameter and $r$ the inter-nuclear separation. The Varshni potential is a short-range repulsive potential energy.

The Hulthén potential is one of the important short-range potentials in physics. Its relevance to diverse areas of physics including nuclear and particle physics, atomic physics, molecular physics, condensed matter and chemical physics has been of great interest and concern to researchers in recent times[26, 27]. The Hulthén potential model takes the form [28]

$$V_H(r) = -\frac{\eta_2 e^{-\delta r}}{1 - e^{-\delta r}}, \qquad (2)$$

where $\delta$ is the screening parameter and $\eta_2$ is the potential strength constant which is sometimes identified with the atomic number when the potential is used for atomic phenomena. Many authors have obtained bound state solutions of SE with this potential. For example, Tazimi and Ghasempour [29] used the NU method to obtain bound state solutions of the three-Dimensional Klein-Gordon equation for two model potentials. Quantization rule was employed by Ikhdair and Abu-Hasna [30] to obtain the solution to the Hulthén potential in arbitrary dimension with a new approximate scheme for the centrifugal term. Okorie *et al* [31] obtain the solution of SE with energy-dependent screened Coulomb potential with the new form of Greene-Aldrich approximation using the NU method.

With the experimental proof of the Schrödinger wave equation, researchers have made great effort to solve the SE by the combination of two or more potentials, which can be used for a wider range of applications [32]. For example, Edet *et al* [33] obtained bound state solutions of the SE for the modified Kratzer potential plus screened Coulomb potential including a centrifugal term. Also, Edet *et al* [34] obtained any $l$-state solutions of the SE interacting with Hellmann-Kratzer potential model. William *et al* [35] obtained bound state solutions of the radial SE by the combination of Hulthén and Hellmann potential within the framework of Nikiforov-Uvarov (NU) method for any arbitrary $\ell$-state, with the Greene-Aldrish approximation in the centrifugal term. Inyang *et al* [36] studied any $\ell$-state solutions of the SE interacting with class of Yukawa-Eckart potentials within the framework of NU method. Hence, motivated by the success of the combination of exponential-type potentials, we seek to investigate the bound state solutions of the SE by the combination of Varshni potential [Eq.(1)] and Hulthen potential [Eq.(2)] for $l \neq 0$ using the NU method. The potential take the form:

$$V(r) = V_V(r) + V_H(r) = \eta_0 - \frac{\eta_0 \eta_1}{r} e^{-\delta r} - \frac{\eta_2 e^{-\delta r}}{1 - e^{-\delta r}} \qquad (3)$$

Considering the achievements in the previous studies, we combine the potentials to allow for more physical application and comparative analysis to existing studies of molecular physics. Also in molecular physics, it is well-known that the potential energy functions with more parameters have a tendency to fit experimental data than those with fewer parameters [37]. The plot of the combine potential with the screening parameter is presented in Fig. 1. However, it must be noted that the exact solution of the SE with the combined potentials in Eq. (3) is not possible due to the presence of the centrifugal term. Therefore, to obtain approximate solutions, we employ a suitable approximation scheme. It is found that such approximation proposed by Greene and Aldrich [8]

$$\frac{1}{r^2} \approx \frac{\delta^2}{\left(1-e^{-\delta r}\right)^2} \tag{4}$$

is a good approximation to the centrifugal or inverse square term for a short range potential which is valid for $\delta \ll 1$.

The paper is organized as follows: In section 2, we derive the bound states solutions of the SE with VHP using NU method and also derive the corresponding normalized wave function. In section 3, we present the results and discussion and finally, we make a concluding remark in section 4.

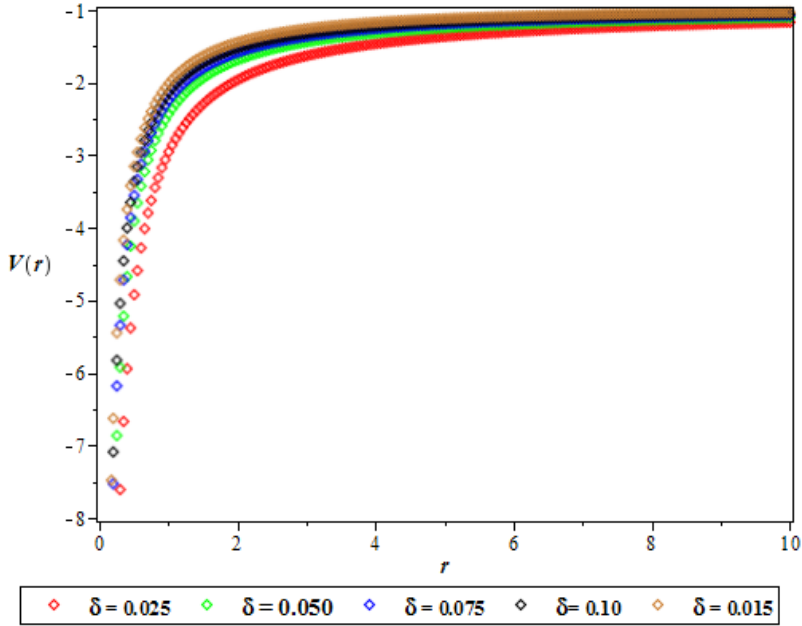

FIGURE 1. Plot of the combined potential as a function of inter-nuclear distance for different screening parameter. We choose $\eta_0 = \eta_1 = -1$, and $\eta_2 = 0.025$

## 2. Bound state solutions of the Schrödinger equation with Varshni plus Hulthen potentials

In this study, we adopt the Nikiforov-Uvarov (NU) method which is based on solving the second-order differential equation of the hypergeometric type. The details can be found in Ref. [38]. The SE takes the form [39]

$$U''_{nl}(r) + \left[\frac{2\mu}{\hbar^2}\left(E_{nl} - V(r)\right) - \frac{(N+2l-1)(N+2l-3)}{4r^2}\right] U_{nl}(r) = 0, \tag{5}$$

where $l$, $\mu$, $r$, $N$ and $\hbar$ are the angular momentum, the reduced mass of the particle, inter-particle distance, the number of space, and reduced plank constant respectively. Substituting Eqs. (3) and (4) into Eq. (5) we obtain

$$U''_{nl}(r) + \left[\frac{2\mu}{\hbar^2}\left(E_{nl} - \eta_0 + \frac{\eta_0\eta_1\delta e^{-\delta r}}{\left(1-e^{-\delta r}\right)} + \frac{\eta_2 e^{-\delta r}}{\left(1-e^{-\delta r}\right)}\right) - \frac{\delta^2(N+2l-1)(N+3l-3)}{4\left(1-e^{-\delta r}\right)^2}\right] U_{nl}(r) = 0. \tag{6}$$

By using coordinate transformation

$$x = e^{-\delta r}, \qquad (7)$$

we obtain the differential equation as

$$U''_{nl}(x) + \frac{1-x}{x(1-x)} U'_{nl}(x) + \frac{1}{x^2(1-x)^2}\left[-(\varepsilon+\beta_0)x^2 + (2\varepsilon+\beta_0)x - (\varepsilon+\gamma)\right]U(x) = 0, \qquad (8)$$

where

$$-\varepsilon = \frac{2\mu E_{nl}}{\hbar^2\delta^2} - \frac{2\eta_0\mu}{\hbar^2\delta^2}, \quad \beta_0 = \frac{2\eta_0\eta_1\mu}{\hbar^2\delta} + \frac{2\eta_2\mu}{\hbar^2\delta^2}, \quad \gamma = \frac{(N+2l-1)(N+2l-3)}{4}. \qquad (9)$$

Now that Eq. (8) and Eq. (1) of Ref. [38] are in the same shape, we have the following parameters:

$$\sigma(x) = -(\varepsilon+\beta_0)x^2 + (2\varepsilon+\beta_0)x - (\varepsilon+\gamma), \ \sigma(x) = x(1-x), \ \tilde{\tau}(x) = 1-x\} \qquad (10)$$

Substituting Eq. (10) into Eq. (11) of Ref. [37], we obtain $\pi(x)$ as

$$\pi(x) = -\frac{x}{2} \pm \sqrt{(A-k)x^2 + (k+B)x + C}, \qquad (11)$$

where

$$A = \frac{1}{4} + \varepsilon + \beta_0, \ B = -(2\varepsilon+\beta_0), \ C = \varepsilon+\gamma\}. \qquad (12)$$

To find the constant $k$, the discriminant of the expression under the square root of Eq. (11) must be equal to zero. As such we have that

$$k = \beta_0 - 2\gamma \pm 2\sqrt{\varepsilon+\gamma}\sqrt{\gamma+\frac{1}{4}}. \qquad (13)$$

Substituting Eq. (13) into Eq. (11) yields $\pi(x)$ as

$$\pi(x) = -\frac{x}{2} \pm \left(\sqrt{\varepsilon+\gamma} + \sqrt{\gamma+\frac{1}{4}}\right)x - \sqrt{\varepsilon+\gamma}, \qquad (14)$$

and $\tau(x)$ can be written as

$$\tau(x) = 1 - 2x - 2\sqrt{\varepsilon+\gamma}\,x - 2\sqrt{\gamma+\frac{1}{4}}\,x + 2\sqrt{\varepsilon+\gamma}. \qquad (15)$$

Taking the derivative of Eq. (15) with respect to $x$, we have

$$\tau'(x) = -2 - 2\left(\sqrt{\varepsilon+\gamma} + \sqrt{\gamma+\frac{1}{4}}\right). \qquad (16)$$

Referring to Eq. (10) of Ref. [38], we define the constant $\lambda$ as,

$$\lambda = -\frac{1}{2} - \sqrt{\varepsilon+\gamma} - \sqrt{\gamma+\frac{1}{4}} + \beta_0 - 2\gamma - 2\sqrt{\varepsilon+\gamma}\sqrt{\gamma+\frac{1}{4}}, \qquad (17)$$

and taking the derivative of $\sigma(x)$ with respect to $x$ from Eq. (10), we have

$$\sigma''(x) = -2. \qquad (18)$$

Substituting Eqs. (15) and (18) into Eq. (13) of Ref. [38], we obtain

$$\lambda_n = n^2 + n + 2n\left(\sqrt{\varepsilon+\gamma} + \sqrt{\gamma+\frac{1}{4}}\right). \qquad (19)$$

By comparing Eqs. (17) and (19), using Eq.(9) yields the energy eigenvalues equation of the VHP as a function of $n$ and $l$ as

$$E_{nl} = \eta_0 + \frac{\delta^2\hbar^2(N+2l-1)(N+2l-3)}{8\mu} - \frac{\delta^2\hbar^2}{8\mu}\left[\frac{\left(n+\frac{1}{2}+\sqrt{\frac{(N+2l-1)(N+2l-3)}{4}+\frac{1}{4}}\right)^2 - Q}{n+\frac{1}{2}+\sqrt{\frac{(N+2l-1)(N+2l-3)}{4}+\frac{1}{4}}}\right]^2 \qquad (20)$$

where

$$Q = \left(\frac{2\eta_0\eta_1\mu}{\hbar^2\delta} + \frac{2\eta_2\mu}{\hbar^2\delta^2}\right) + \frac{(N+2l-1)(N+2l-3)}{4} \qquad (21)$$

To obtain the corresponding wavefunction, we consider Eq. (3) of Ref [38], and upon substituting Eqs. (10) and (14) and integrating, we get

$$\phi(x) = x^{\sqrt{\varepsilon+\gamma}}(1-x)^{\frac{1}{2}+\sqrt{\frac{1}{4}+\gamma}} \qquad (22)$$

To get the hypergeometric function, we first determine the weight function $\rho(x)$. By substituting Eqs. (10) and (14) into Eq. (3) of Ref. [38] and after integrating, we obtain

$$\rho(x) = x^{2\sqrt{\varepsilon+\gamma}}(1-x)^{2\sqrt{\frac{1}{4}+\gamma}}. \qquad (23)$$

Hence, by substituting Eqs. (10) and (23) into Eq. (2) yields the Rodrigues equation given as

$$y(x) = N_{nl}x^{-2\sqrt{\varepsilon+\gamma}}(1-x)^{-2\sqrt{\frac{1}{4}+\gamma}}\frac{d^n}{dx^n}\left[x^{n+2\sqrt{\varepsilon+\gamma}}(1-x)^{n+2\sqrt{\frac{1}{4}+\gamma}}\right], \qquad (24)$$

where $N_{nl}$ is the normalization constant. Equation (24) is equivalent to

$$P_n^{\left(2\sqrt{\varepsilon+\gamma},\,2\sqrt{\frac{1}{4}+\gamma}\right)}(1-2x) \tag{25}$$

where $p_n^{(\alpha,\beta)}$ is the Jacobi Polynomials. Hence, the wave function is given as

$$\psi_{nl}(x) = N_{nl} x^{\sqrt{\varepsilon+\gamma}} (1-x)^{\frac{1}{2}+\sqrt{\frac{1}{4}+\gamma}} P_n^{\left(2\sqrt{\varepsilon+\gamma},\,2\sqrt{\frac{1}{4}+\gamma}\right)}(1-2x). \tag{26}$$

Using the normalization condition, we obtain the normalization constant as follows:

$$\int_0^\infty |\psi_{nl}(r)|^2 \, dr = 1 \tag{27}$$

From our coordinate transformation of Eq. (7), we have that

$$-\frac{1}{\delta x} \int_1^0 |\psi_{nl}|^2 \, dx = 1 \tag{28}$$

By letting, $y = 1 - 2x$, we have

$$\frac{N_{nl}^2}{\delta} \int_{-1}^{1} \left(\frac{1-y}{2}\right)^{2\sqrt{\varepsilon+\gamma}} \left(\frac{1+y}{2}\right)^{1+2\sqrt{\frac{1}{4}+\gamma}} \left[ P_n^{\left(2\sqrt{\varepsilon+\gamma},\,2\sqrt{\frac{1}{4}+\gamma}\right)} y \right]^2 dy = 1. \tag{29}$$

Let

$$\upsilon = 1 + 2\sqrt{\frac{1}{4}+\gamma},\ \text{and}\ \mu - 1 = 2\sqrt{\frac{1}{4}+\gamma},\ u = 2\sqrt{\varepsilon+\gamma}. \tag{30}$$

By substituting Eq. (29) into Eq. (28), using Eq. (30), we have

$$\frac{N_{nl}^2}{\delta} \int_{-1}^{1} \left(\frac{1-y}{2}\right)^u \left(\frac{1+y}{2}\right)^\upsilon \left[ P_n^{(2u,\,\upsilon-1)} y \right]^2 dy = 1. \tag{31}$$

According to Onate and Ojonubah [40], integral of the form in Eq. (31) can be expressed as

$$\int_{-1}^{1} \left(\frac{1-p}{2}\right)^x \left(\frac{1+p}{2}\right)^y \left[ P_n^{(2x,2y-1)} p \right]^2 dp = \frac{2\Gamma(x+n+1)\Gamma(y+n+1)}{n! x \Gamma(x+y+n+1)} \tag{32}$$

Hence, by comparing Eq. (31) with the standard integral of Eq. (32), we obtain the normalization constant as

$$N_{nl} = \sqrt{\frac{n! u \delta \Gamma(u+\upsilon+n+1)}{2\Gamma(u+n+1)\Gamma(\upsilon+n+1)}}. \tag{33}$$

**3. Discussion**

Solution of the radial N-dimensional SE for the newly proposed potential obtained by the superposition of Varshni and Hulthen potentials otherwise known as Varshni-Hulthen potential (VHP) are obtain within the framework of NU method by approximation to the centrifugal barrier. In Table 1, we reported the numerical energy eigenvalues $(eV)$ with $\hbar = 2\mu = 1$ for 1S, 2S, 2P, 3S, 3P, 3d, 4S and 4P states with the potential strength: $\eta_0 = \eta_1 = -1$, $\eta_2 = 0.025$ and for $N = 3, 4,$ and $5,$ respectively. We observed that the energy increases as the screening parameters and $N$ increases. In Table 2, we have presented the energy eigenvalues $(eV)$ with $\hbar = 2\mu = 1$ of the VHP for $N = 3$ with three different values of the potential range: $\eta_0 = -1, \ \eta_1 = -2, \ \eta_2 = 0.050; \eta_0 = -2, \ \eta_1 = -1, \ \eta_2 = 0.075$ and $\eta_0 = \eta_1 = -2,$ $\eta_2 = 0.10,$ respectively, and observed that the energy increases as the screening parameters increases for constant $N$. We have plotted the graph of energy eigenvalues against the number of dimensional space, potential strength, screening parameter, and the reduced mass in the ground and excited states. In Fig 2, we show the variation of energy as a function of $N$ and observed that the energy increases as both $N$ and $\delta$ increases. From Figs. 3(a) and (b) – Figs. 5(a) and (b) respectively, we plotted the ground and excited states energy eigenvalues of the different quantum states as a function of the VHP strengths. We observed that there is a decrease in energy in both the ground and excited states as the potential strength, $\eta_0, \ \eta_1,$ and $\eta_2$ , respectively, increases.

TABLE 1. Energy eigenvalues $(eV)$ of the Varshni-Hulthen potential for $\eta_0 = \eta_1 = -1, \ \eta_2 = 0.025$ with $\hbar = 2\mu = 1$

| State | $\delta$ | $N = 3$ | $N = 4$ | $N = 5$ |
|---|---|---|---|---|
| 1s | 0.025 | -1.975781250 | -1.419947195 | -1.224992578 |
|    | 0.050 | -1.526875000 | -1.213252345 | -1.101834379 |
|    | 0.075 | -1.397725694 | -1.148269303 | -1.057252171 |
|    | 0.100 | -1.333125000 | -1.111565340 | -1.027444000 |
|    | 0.150 | -1.262152778 | -1.062800124 | -0.9783441671 |
| 2s | 0.025 | -1.226250000 | -1.136131084 | -1.086889280 |
|    | 0.050 | -1.106875000 | -1.055772344 | -1.026860418 |
|    | 0.075 | -1.068611111 | -1.027534543 | -1.002633126 |
|    | 0.100 | -1.047656250 | -1.010587522 | -0.9858640000 |
|    | 0.150 | -1.023819444 | -0.9889295445 | -0.9597536576 |
| 2p | 0.025 | -1.224992578 | -1.134249170 | -1.084383117 |
|    | 0.050 | -1.101834379 | -1.048236728 | -1.016831266 |
|    | 0.075 | -1.057252171 | -1.010572100 | -0.9800715659 |
|    | 0.100 | -1.027444000 | -0.9804380625 | -0.9457871111 |
|    | 0.150 | -0.9783441671 | -0.9212500704 | -0.8699012826 |
| 3s | 0.025 | -1.088142361 | -1.058702155 | -1.039373438 |
|    | 0.050 | -1.031875000 | -1.015446221 | -1.003900782 |
|    | 0.075 | -1.013913966 | -1.001158640 | -0.9908992733 |
|    | 0.100 | -1.005902778 | -0.9950120013 | -0.9844360000 |
|    | 0.150 | -1.004683642 | -0.9961284666 | -0.9827782293 |
| 3p | 0.025 | -1.086889280 | -1.056823915 | -1.036870313 |
|    | 0.050 | -1.026860418 | -1.007932642 | -0.9938898525 |
|    | 0.075 | -1.002633126 | -0.9842622719 | -0.9683923310 |
|    | 0.100 | -0.9858640000 | -0.9650092156 | -0.9444802500 |
|    | 0.150 | -0.9597536576 | -0.9289088370 | -0.8933042371 |
| 3d | 0.025 | -1.084383117 | -1.053693514 | -1.033115626 |
|    | 0.050 | -1.016831266 | -0.9954100212 | -0.9788734727 |
|    | 0.075 | -0.9800715659 | -0.9561018046 | -0.9346320845 |
|    | 0.100 | -0.9457871111 | -0.9150053890 | -0.8845475625 |

| State | δ | | | |
|---|---|---|---|---|
| | 0.150 | −0.8699012826 | −0.8168854194 | −0.7591039275 |
| 4s | 0.025 | −1.040625000 | −1.027702596 | −1.018280391 |
| | 0.050 | −1.008906250 | −1.002307899 | −0.9968737506 |
| | 0.075 | −1.002152778 | −0.9980821361 | −0.9935681755 |
| | 0.100 | −1.004414062 | −1.002429482 | −0.9981750400 |
| | 0.150 | −1.027517361 | −1.030202097 | −1.025835317 |
| 4p | 0.025 | −1.039373438 | −1.025825868 | −1.015778672 |
| | 0.050 | −1.003900782 | −0.9948033881 | −0.9868712556 |
| | 0.075 | −0.9908992733 | −0.9812129600 | −0.9710865137 |
| | 0.100 | −0.9844360000 | −0.9724870563 | −0.9582753600 |
| | 0.150 | −0.9827782293 | −0.9631717038 | −0.9365364617 |

TABLE 2. Energy eigenvalues $(eV)$ of the Varshni-Hulthen potential for $N=3$ with $\hbar=2\mu=1$

| State | δ | $\eta_0=-1,\ \eta_1=-2,\ \eta_2=0.050$ | $\eta_0=-2,\ \eta_1=-1,\ \eta_2=0.075$ | $\eta_0=\eta_1=-2,\ \eta_2=0.10$ |
|---|---|---|---|---|
| 1s | 0.025 | −4.951406250 | −8.189531250 | −17.90265625 |
| | 0.050 | −3.178125000 | −4.979375000 | −10.85562500 |
| | 0.075 | −2.682934026 | −4.144531249 | −8.920017362 |
| | 0.100 | −2.445000000 | −3.763125000 | −8.012500000 |
| | 0.150 | −2.199236110 | −3.391874999 | −7.115069440 |
| 2s | 0.025 | −1.951875000 | −3.502500000 | −5.903125000 |
| | 0.050 | −1.492500000 | −2.684375000 | −4.107500000 |
| | 0.075 | −1.353819444 | −2.461250000 | −3.590902778 |
| | 0.100 | −1.280625000 | −2.352656250 | −3.332500000 |
| | 0.150 | −1.195277778 | −2.236875000 | −3.048611110 |
| 2p | 0.025 | −1.950609766 | −3.501230860 | −5.901844140 |
| | 0.050 | −1.487412504 | −2.679271879 | −4.102318752 |
| | 0.075 | −1.342319880 | −2.449715279 | −3.579121962 |
| | 0.100 | −1.260100250 | −2.332069000 | −3.311350250 |
| | 0.150 | −1.148818125 | −2.190274722 | −3.000182708 |
| 3s | 0.025 | −1.397100694 | −2.635225694 | −3.681684028 |
| | 0.050 | −1.183125000 | −2.262152778 | −2.860625000 |
| | 0.075 | −1.113937114 | −2.155781250 | −2.610279707 |
| | 0.100 | −1.076111111 | −2.102569444 | −2.476944444 |
| | 0.150 | −1.034359568 | −2.047986111 | −2.320563272 |
| 3p | 0.025 | −1.395844140 | −2.633967404 | −3.680420530 |
| | 0.050 | −1.178089585 | −2.257110418 | −2.855547918 |
| | 0.075 | −1.102593775 | −2.144422285 | −2.598811367 |
| | 0.100 | −1.055933444 | −2.082364000 | −2.456489000 |
| | 0.150 | −0.9889920835 | −2.002556127 | −2.274320787 |
| 3d | 0.025 | −1.393331033 | −2.631450825 | −3.677893532 |
| | 0.050 | −1.168018766 | −2.247025710 | −2.845393765 |
| | 0.075 | −1.079907214 | −2.121704475 | −2.575874807 |
| | 0.100 | −1.015578778 | −2.041953778 | −2.415578778 |
| | 0.150 | −0.8982647085 | −1.911703752 | −2.181843412 |
| 4s | 0.025 | −1.203750000 | −2.332500000 | −2.905000000 |
| | 0.050 | −1.078125000 | −2.117656250 | −2.427500000 |
| | 0.075 | −1.037361111 | −2.056250000 | −2.274444444 |
| | 0.100 | −1.017656250 | −2.028164062 | −2.190625000 |
| | 0.150 | −1.007569444 | −2.011406250 | −2.095277778 |
| 4p | 0.025 | −1.202496484 | −2.331245508 | −2.903742578 |
| | 0.050 | −1.073107813 | −2.112635157 | −2.422459376 |

| | | | |
|---|---|---|---|
| 0.075 | −1.026072450 | −2.044952550 | −2.263085471 |
| 0.100 | −0.9976000625 | −2.008092250 | −2.170412562 |
| 0.150 | −0.9625842189 | −1.966385868 | −2.049800365 |

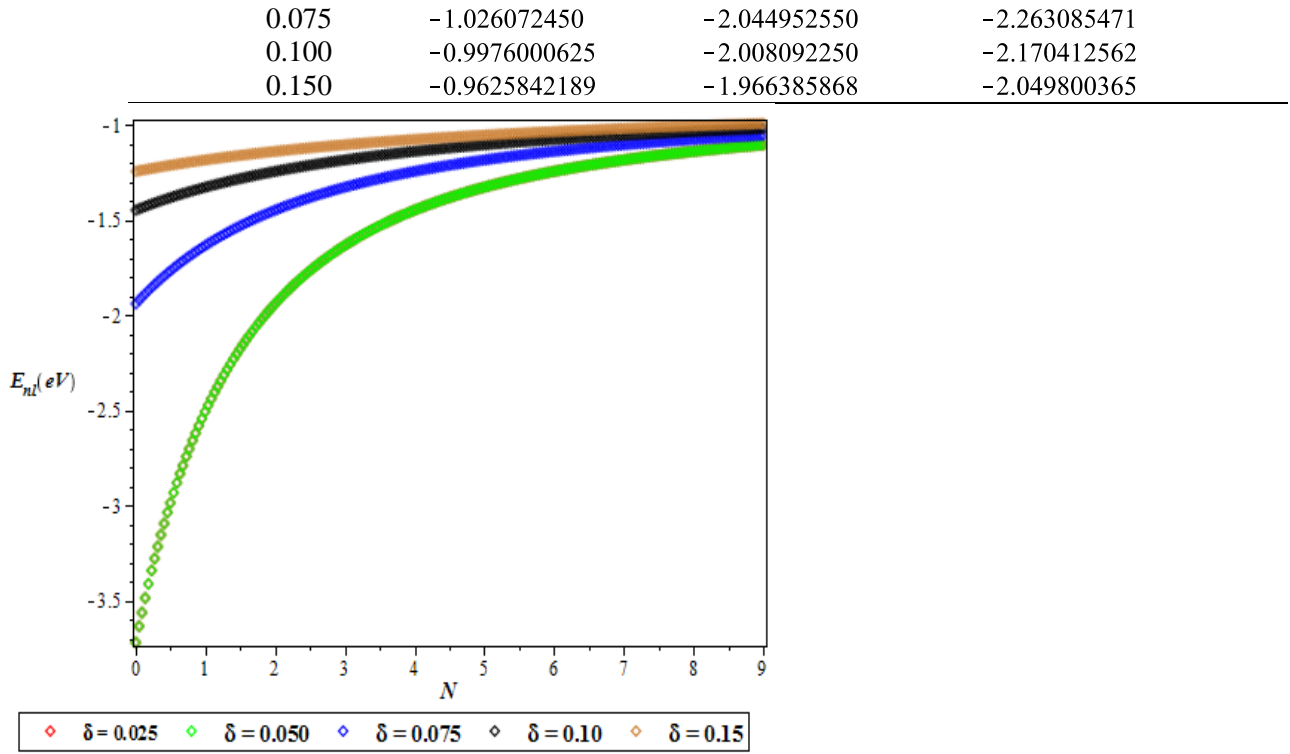

FIGURE 2. Variation of energy eigenvalues as function of N for different $l$ and $\delta$. We choose $a = b = -1$, and $c = 0.025$

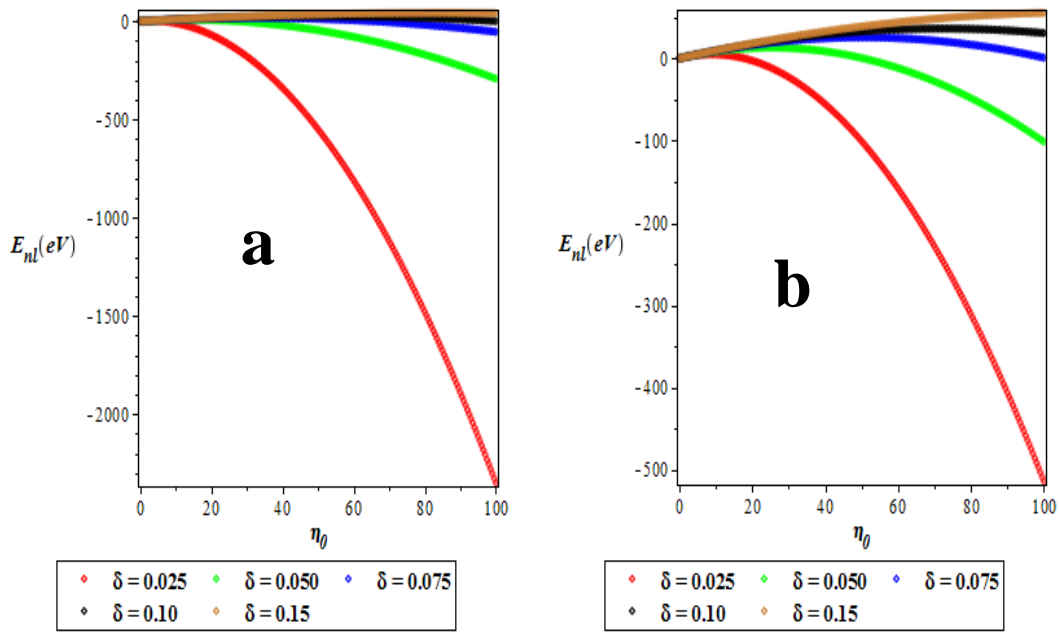

FIGURE 3(a). The plot of the ground state energy spectra for various $l$ as a function of $\eta_0$. (b). Variation of the first excited state energy spectra for different $l$ as a function of $\eta_0$. We choose $a = b = -1$, $c = 0.025$, and $\delta = 0.025$ for the ground and excited states.

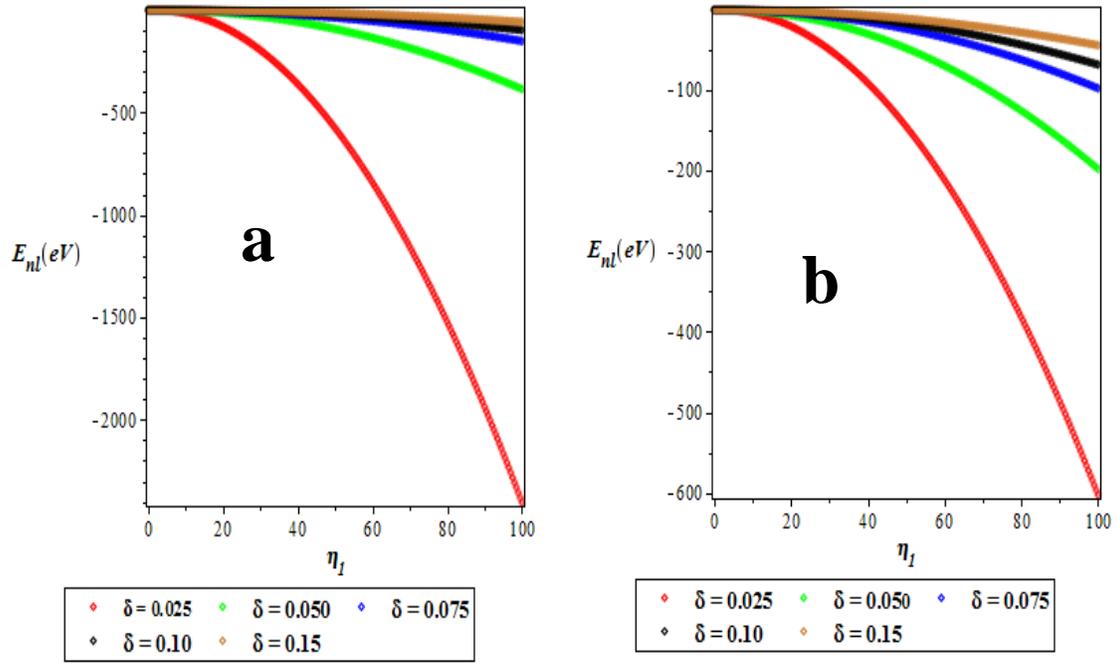

FIGURE 4(a). Variation of the ground state energy spectra for various $l$ as a function of $\eta_1$. (b) A plot of the first excited state energy spectra for various $l$ as a function of $\eta_1$. We choose $a = b = -1$, $c = 0.025$, and $\delta = 0.025$ for the ground and excited states.

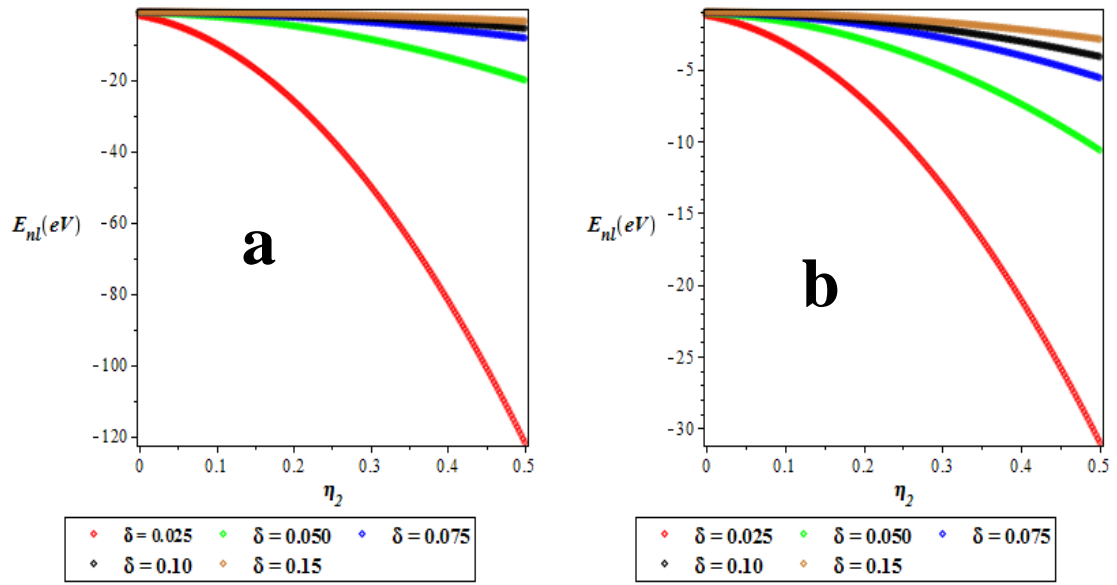

FIGURE 5(a). Variation of the ground state energy spectra for various $l$ as a function of $\eta_2$. (b). The plot of the first excited state energy spectra for various $l$ as a function of $\eta_2$. We choose $a = b = -1$, $c = 0.025$, and $\delta = 0.025$ for the ground and excited states.

The variation of energy eigenvalues of VHP as a function of the screening parameter is as shown in Fig. 6(a) and 6(b). We observed that the energy increases as the screening parameter increases in the ground and excited states. In Fig. 7(a) and 7(b), we show the variation of energy eigenvalues as a function of the reduced mass. It is observed that there is a decrease in the ground and excited states energy for different quantum states as the reduced mass, $\mu$ increases.

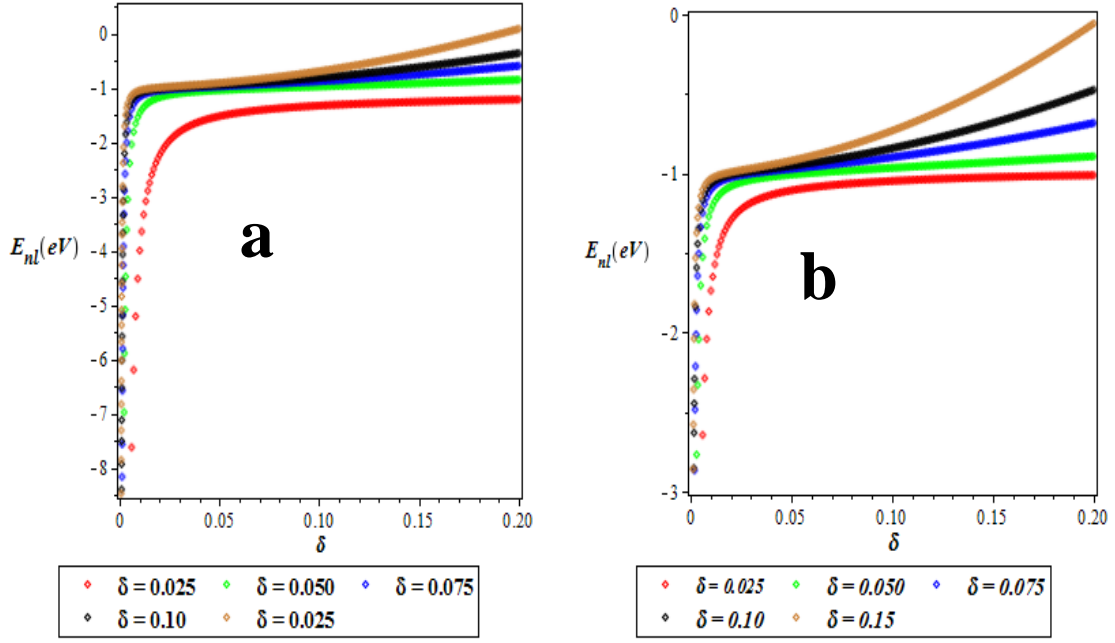

FIGURE 6(a). Variation of the ground state energy spectrum for various $l$ as a function of the screening parameter $(\delta)$ (b). A plot of the first excited state energy spectrum for different $l$ as a function of the screening parameter $(\delta)$. We choose $a = b = -1$, $c = 0.025$, and $\delta = 0.025$ for the ground and excited states

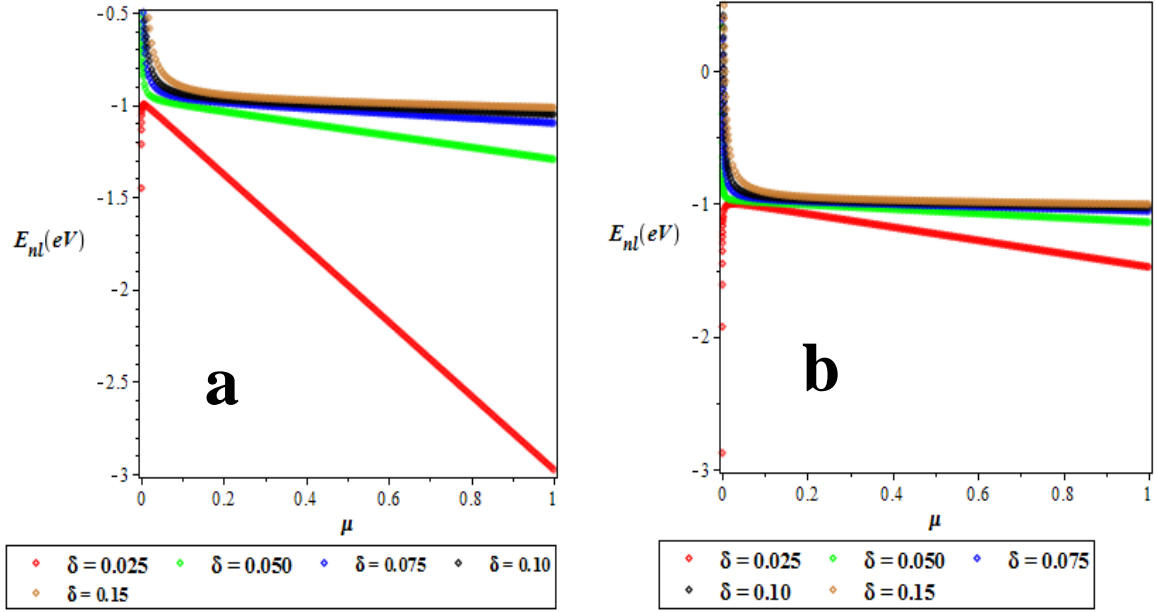

FIGURE 7(a). Variation of the ground state energy spectra for various $l$ as a function of the reduced mass $(\mu)$ (b). The plot of the first excited state energy spectra for different $l$ as a function of the reduced mass $(\mu)$. We choose $a = b = -1$, $c = 0.025$, and $\delta = 0.025$ for the ground and excited states.

**Special cases**

1. When we set $\eta_2 = 0$ in Eq. (20), we obtain the energy eigenvalues for Varshni potential

$$E_{nl} = \eta_0 + \frac{\delta^2\hbar^2(N+2l-1)(N+2l-3)}{8\mu} - \frac{\delta^2\hbar^2}{8\mu}\left[\frac{\left(n+\frac{1}{2}+\sqrt{\frac{(N+2l-1)(N+2l-3)}{4}+\frac{1}{4}}\right)^2 - P}{n+\frac{1}{2}+\sqrt{\frac{(N+2l-1)(N+2l-3)}{4}+\frac{1}{4}}}\right]^2 \quad (34)$$

where

$$P = \frac{2\eta_0\eta_1\mu}{\hbar^2\delta} + \frac{(N+2l-1)(N+2l-3)}{4} \quad (35)$$

The numerical energy eigenvalues of Eqs. (34) and (35) for $N = 3$ as presented in Table 3, were computed for three different potential strength for 1s, 2s, 2p, 3s, 3p, 3d, 4s, 4p, 4d and 4f states and were found to be in conformity when compared with the work of Ebonwonyi *et al.* [41] using formula method.

2. If we set $\eta_0 = \eta_1 = 0$ in Eq. (20), we obtain the energy eigenvalue equation for Hulthen potential

$$E_{nl} = \frac{\delta^2\hbar^2(N+2l-1)(N+2l-3)}{8\mu} - \frac{\delta^2\hbar^2}{8\mu}\left[\frac{\left(n+\frac{1}{2}+\sqrt{\frac{(N+2l-1)(N+2l-3)}{4}+\frac{1}{4}}\right)^2 - R}{n+\frac{1}{2}+\sqrt{\frac{(N+2l-1)(N+2l-3)}{4}+\frac{1}{4}}}\right]^2 \quad (36)$$

where

$$R = \frac{2\eta_2\mu}{\hbar^2\delta^2} + \frac{(N+2l-1)(N+2l-3)}{4} \quad (37)$$

The numerical energy eigenvalues of Hulthen potential presented in Eqs. (36) and (37) for $N = 3$ were computed for 2p, 3p, 3d, 4p, 4d and 4f states and was compare with the work of Qiang *et al.* [42] using EQR, Bayrak *et al.* [43] using AIM, and Ikhdair [44] using numerical method and were found to be in excellent agreement as presented in Table 4.

3 If we set $N = l = 0$ in Eq. (20), we obtain the energy eigenvalue equation for the s-wave of VHP as

$$E_{nl} = \eta_0 - \frac{\delta^2\hbar^2}{8\mu}\left[\frac{\left(n+\frac{1}{2}\right)^2 - \left(\frac{2\eta_0\eta_1\mu}{\hbar^2\delta} + \frac{2\eta_2\mu}{\hbar^2\delta^2}\right)}{n+\frac{1}{2}}\right]^2 \quad (38)$$

TABLE 3. Energy eigenvalues $(eV)$ of Varshni potential as a function of the screening parameter $\delta$ for in atomic mass units $(\hbar = \mu = 1)$

| State | $\delta$ | Present method $a=b=-1$ | Present method $a=-1, b=-2$ | Present method $a=-2, b=-1$ | (FM) [40] $a=b=-1$ |
|---|---|---|---|---|---|
| 1S | 0.001 | −1.063124562 | −2.001000250 | −3.001000250 | |
|  | 0.050 | −1.092656250 | −2.050625000 | −3.050625000 | |
|  | 0.100 | −1.120625000 | −2.102500000 | −3.102500000 | |
| 2S | 0.001 | −1.063001000 | −1.251001000 | −2.251001000 | |
|  | 0.050 | −1.090000000 | −1.302500000 | −2.302500000 | |
|  | 0.100 | −1.122500000 | −1.360000000 | −2.360000000 | |
| 2P | 0.001 | −1.063124562 | −1.251249562 | −2.251249562 | −1.0617502 |
|  | 0.050 | −1.092656250 | −1.311406250 | −2.311406250 | −1.0256250 |
|  | 0.100 | −1.120625000 | −1.370625000 | −2.370625000 | −0.9900000 |
| 3S | 0.001 | −1.028280028 | −1.112113361 | −2.112113361 | |
|  | 0.050 | −1.058402778 | −1.166736111 | −2.166736111 | |
|  | 0.100 | −1.100277778 | −1.233611111 | −2.233611111 | |
| 3p | 0.001 | −1.028334111 | −1.112223000 | −2.112223000 | |
|  | 0.050 | −1.057500000 | −1.168611111 | −2.168611111 | |
|  | 0.100 | −1.091111111 | −1.230000000 | −2.230000000 | − |
| 3d | 0.001 | −1.028386250 | −1.112330694 | −2.112330694 | −1.0269447 |
|  | 0.050 | −1.051736111 | −1.165625000 | −2.165625000 | −0.9867361 |
|  | 0.100 | −1.062500000 | −1.206944444 | −2.206944444 | −0.9469444 |
| 4S | 0.001 | −1.016129000 | −1.063504000 | −2.063504000 | − |
|  | 0.050 | −1.050625000 | −1.122500000 | −2.122500000 | − |
|  | 0.100 | −1.105625000 | −1.202500000 | −2.202500000 | − |
| 4p | 0.001 | −1.016158766 | −1.063565016 | −2.063565016 | −1.0150656 |
|  | 0.050 | −1.048476562 | −1.121914062 | −2.121914062 | −0.9951563 |
|  | 0.100 | −1.093906250 | −1.193906250 | −2.193906250 | −0.9900000 |
| 4d | 0.001 | −1.028386250 | −1.063624062 | −2.063624062 | −1.0149391 |
|  | 0.050 | −1.051736111 | −1.116406250 | −2.116406250 | −0.9851563 |
|  | 0.100 | −1.062500000 | −1.063624062 | −2.165625000 | −0.9625000 |
| 4f | 0.001 | −1.016212391 | −1.063681141 | −2.063681141 | −1.0147502 |
|  | 0.050 | −1.029414062 | −1.105976562 | −2.105976562 | −0.9725000 |
|  | 0.100 | −1.011406250 | −1.117656250 | −2.117656250 | −0.9306250 |

TABLE 4. Energy eigenvalues $(eV)$ of Hulthén potential as a function of the screening parameter for $N=3$ in atomic mass units $(\hbar = \mu = 1)$

| State | $\delta$ | Present method | EQR[42] | AIM[43] | Numerical [44] |
|---|---|---|---|---|---|
| 2p | 0.025 | −0.1127611 | −0.1128125 | −0.1128125 | −0.112760 |
|  | 0.050 | −0.1010442 | −0.1012500 | −0.1012500 | −0.101042 |
|  | 0.075 | −0.0898495 | −0.0903125 | −0.0903125 | −0.089847 |
|     | 0.100 | - 0.0791769 | - 0.0800000 | - 0.0800000 | - 0.079179 |
|     | 0.150 | - 0.0593981 | - 0.0612500 | - 0.0612500 | - 0.059441 |
| 3p  | 0.025 | - 0.0437072 | - 0.0437590 | - 0.0437590 | - 0.043706 |
|     | 0.050 | - 0.0331623 | - 0.0333681 | - 0.033368  | - 0.033164 |
|     | 0.075 | - 0.0239207 | - 0.0243837 | - 0.0243837 | - 0.023939 |
|     | 0.100 | - 0.0159825 | - 0.0168056 | - 0.0168056 | - 0.016053 |
|     | 0.150 | - 0.0040162 | - 0.0058681 | - 0.0058681 | - 0.004466 |
| 3d  | 0.025 | - 0.0436044 | - 0.0437587 | - 0.0437587 | - 0.043603 |
|     | 0.050 | - 0.0327508 | - 0.0333681 | - 0.0333681 | - 0.032753 |
|     | 0.075 | - 0.0229948 | - 0.0243837 | - 0.0243837 | - 0.023030 |
|     | 0.100 | - 0.0143364 | - 0.0168055 | - 0.0168055 | - 0.014484 |
|     | 0.150 | - 0.0003124 | - 0.0058681 | - 0.0058681 | - 0.001396 |
| 4p  | 0.025 | - 0.0199486 | - 0.0200000 | - 0.0200000 | - 0.019948 |
|     | 0.050 | - 0.0110442 | - 0.0112500 | - 0.0112500 | - 0.011058 |
|     | 0.075 | - 0.0045370 | - 0.0050000 | - 0.0050000 | - 0.004621 |
|     | 0.100 | - 0.0004269 | - 0.0012500 | - 0.0012500 | - 0.000755 |
| 4d  | 0.025 | - 0.0198457 | - 0.0200000 | - 0.0200000 | - 0.019846 |
|     | 0.050 | - 0.0106327 | - 0.0112500 | - 0.0112500 | - 0.010667 |
|     | 0.075 | - 0.0036111 | - 0.0050000 | - 0.0050000 | - 0.003834 |
| 4f  | 0.025 | - 0.0196914 | - 0.0200000 | - 0.0200000 | - 0.019691 |
|     | 0.050 | - 0.0100154 | - 0.0112500 | - 0.0112500 | - 0.010062 |
|     | 0.075 | - 0.0022222 | - 0.0050000 | - 0.0050000 | - 0.002556 |

## 4     Conclusion

The analytical solutions of the N-dimensional Schrödinger equation for the newly proposed Varshni-Hulthen potential are obtained via NU method by using Greene-Aldrich approximation scheme to the centrifugal barrier. The numerical energy eigenvalues and the corresponding normalized eigenfunctions are obtained for various values of $l$ and $N$. We have also obtained the numerical energy eigenvalues for two special cases of the newly proposed potential and their results were found to be in agreement with the existing literature. However, the behavior of the energy for the ground state and several excited states is illustrated graphically. Therefore, studying of analytical solution of the N-dimensional Schrödinger equation for the newly proposed Varshni-Hulthen potential could provide valuable information on the quantum mechanics dynamics at atomic and molecular physics and opens new window.

**References**

[1]     O. Aysel *Approximate bound state solutions of the Hellmann plus Kratzer potential in N-dimensional space*, GU J. Sci. 33 (2020) 3 780-793. https://doi.org/10.35378/gujs.672684

[2]     W.A. Yahya and K.J. Oyewumi, *Thermodynamic properties and approximate solutions of the l-state Poschi-Teller-type potential*, J. of the Association of Arab Univ. for basic and applied sciences 21(2016) 1 53-58. http://doi.org/10.1016/j.jaubas.2015.04.001

[3]     K. J. Oyewumi, O. J. Oluwadare, *The scattering phase shifts of the Hulth´en-type potential plus Yukawa potential*, Eur. Phys. J. Plus, 131 (2016) 295, 1-10, https://doi.org/10.1140/epjp/i2016-16295-y

[4]     A.N. Ikot et al*., Bound state solutions of the Schrödinger equation with energy dependent molecular Kratzer potential via asymptotic iteration method*. Eclet. Quim. J. 45 (2020) 65. https://doi.org/10.26850/1678-4618eqj. v45.1.p65-76 5.

[5] E. S. William et al., *Analytical investigation of the single-particle energy spectrum in magic nuclei of $^{56}Ni$ and $^{116}Sn$*. European J. of applied physics 2 (2020) 6. https://doi.org/10.24018/ejphysics.2020.2.6.28

[6] H. Louis, B. I. Ita, and N. I. Nzeata, *Approximate solution of the Schrödinger equation with Manning-Rosen plus Hellmann potential and its thermodynamic properties using the proper quantization rule*. The European Physical Journal Plus, 134 (2019) 7, 1-13. https://doi.org/10.1140/epjp/i2019-12835-3

[7] J. Lu, *Analytic Quantum Mechanics of Diatomic Molecules with Empirical Potentials*, Phys. Scr. 72 (2005) 5, 349-352. https://doi.org/10.1238/Physica.Regular.072a00349

[8] R. L. Greene, C. Aldrich, *Variational wave functions for a screened Coulomb potential*, Phys. Rev. A, 14 (1976) 6, 2363–2366. https://doi.org/10.1103/physreva.14.2363

[9] C.S. Jia, T. Chen, L.-G. Cui, *Approximate analytical solutions of the Dirac equation with the generalized Pöschl–Teller potential including the pseudo-centrifugal term*, Phys. Lett. A 373, (2009) 18-19, 1621-1626. https://doi.org/10.1016/j.physleta.2009.03.006

[10] E. H. Hill, *The Theory of Vector Spherical Harmonics*, Am. J. Phys. 22 (1954) 4 210-0. https://doi.org/10.1119/1.1933682

[11] C. L. Pekeris. *The Rotation-Vibration Coupling in Diatomic Molecules*, Phys. Rev. 45 (1934) 2, 98-103. https://doi.org/10.1103/PhysRev.45.98

[12] B.H. Yazarloo, H. Hassanabadi, S. Zarinkarmar, *Oscillator strengths based on the Möbius square potential under Schrödinger equation*, Eur. Phys. J. Plus 127, (2012) 5, 51-471. https://doi.org/10.1140/epjp/i2012-12051-9

[13] W. C. Qiang and S. H. Dong. Analytical approximations to the solutions of the Manning–Rosen potential with centrifugal term. Phys. lett. A 368 (2007) 1-2, 13–17. doi:10.1016/j.physleta.2007.03.057

[14] S. H. Dong, W. C. Qiang, G. H. Sun and V. B. Bezerra. *Analytical approximations to the l-wave solutions of the Schrödinger equation with the Eckart potential.* J. Phys. A Math. Theor. 40 (2007) 34, 10535–10540. doi:10.1088/1751-8113/40/34/010

[15] G. F. Wei, C. Y. Long and S. H. Dong. *The scattering of the Manning–Rosen potential with centrifugal term.* Phys. Lett. A 372 (2008) 2592–2596. doi:10.1016/j.physleta.2007.12.042

[16] S. H. Dong, X. Y. Gu. *Arbitrary l-state solutions of the Schrödinger equation with the Deng-Fan molecular potential*. J. Phys: Conf. 96 (2008) 012109, 1-7. doi:10.1088/1742-6596/96/1/012109

[17] S. H. Dong, W. C. Qiang, J. García-Ravelo. *Analytical approximations to the Schrödinger equation for a second Pöschl–Teller-like potential with centrifugal term*. Intl J. Mod. Phys. A, 23 (2008) 10, 1537–1544. doi:10.1142/S0217751X0803944X

[18] J. E. Ntibi, E. P. Inyang, E. P. Inyang and E. S. William, *Relativistic treatment of D-dimensional Klein-Gordon equation with Yukawa potential*, Intl. J. Innov sci, engr. Tech. 7 (2020) 11. https//doi/10.13140/RG.2.2.32473.34406

[19] K.J. Oyewumi, F.O. Akinpelu, and A.D. Agboola, *Exactly Complete Solutions of the Pseudoharmonic Potential in N-Dimensions*, Int J. Theor. Phys. 47 (2008) 4, 1039-1057. https://doi.org/10.1007/s10773-007-9532-x

[20] B. Gönül and M. *Koçak Explicit solutions for N-dimensional Schrödinger equations with position-dependent mass* J. Maths Phys. 47, 102101 (2006) 10, 1-7. https://doi.org/10.1063/1.2354333

[21]   K.J. Oyewumi, *Analytical solutions of the kratzer-fues potential in an arbitrary number of dimensions*, Found. Phys. Lett. 18 (2005) 1, 75-84. https://doi.org/10.1007/s10702-005-2481-9

[22]   S.M. Ikhdair and R. Sever, *exact bound states of the d-dimensional Klein–Gordon equation with equal scalar and vector ring-shaped pseudoharmonic potential*, Int. J. Mod. Phys. C 19 (2008) 9, 1425-1442. https://doi.org/10.1142/S0129183108012923

[23]   S. H. Dong. Wave Equations in Higher Dimensions. (Springer, 2011) 1-299. doi:10.1007/978-94-007-1917-0

[24]   Y.P. Varshni, *Comparative Study of Potential Energy Functions for Diatomic Molecules*, Rev. Mod. Phys. 29 (1957) 4, 664-682. https://doi.org/10.1103/RevModPhys.29.664

[25]   E. P. Inyang, E. P. Inyang, E. S. William and E. E. Ibekwe, *Study on the applicability of Varshn potential to predict the mass spectra of quark-antiquark systems in non-relativistic framework*, Jord. J. Phys. (2020) JJ14-11-10-020

[26]   C. O. Edet, P.O. Okoi, *Any l-state solutions of the Schrodinger equation for q-deformed Hulthen plus generalized inverse quadratic Yukawa potential in arbitrary dimensions*, Rev Mex Fis, 65 (2019) 333-344. https://doi.org/10.31349/RevMexFis.65.333

[27]   I. B. Okon, O. Popoola, E. E. Ituen, *Bound state solution to Schrodinger equation with Hulthen plus exponential Coulombic potential with centrifugal potential barrier using parametric NikiforovUvarov method*, Intl J. Rec. adv. Phys. ,5 (2016) 2. 1-15. https://doi.org/10.14810/ijrap.2016.5101 1

[28]   O. Bayrak, and I Boztosun, *Bound state solutions of the Hulthén potential by using the asymptotic iteration method. Physica Scripta, 76* (2007) 1, 92–96. https://doi.org/10.1088/0031-8949/76/1/016

[29]   Tazimi and A. Ghasempour, *Bound State Solutions of Three-Dimensional Klein-Gordon Equation for Two Model Potentials by NU Method*, N. Advances in High Energy Physics (2020) 2541837, 1-10. https://doi.org/10.1155/2020/2541837

[30]   S. M. Ikhdair and J. Abu-Hasna, *Quantization rule solution to the Hulthén potential in arbitrary dimension with a new approximate scheme for the centrifugal term. Physica Scripta, 83* J. (2011) 2, *025002*, 1-8. https://doi.org/10.1088/0031-8949/83/02/025002

[31]   U.S. Okorie, A. N. Ikot, P.O. Amadi, A. T. Ngiangia and E. E. Ibekwe *Approximate solutions of the Schrodinger equation with energy- dependent screened Coulomb potential in D-dimensions*. Ecletica Quimica J. 45 (2020) 4. https://doi.org/10.26850/1678-4618eqj.v45.4.2020.p40-56

[32]   E. P. Inyang, E. P. Inyang, I. O. Akpan, J. E. Ntibi and E. S. William, *Analytical solutions of the Schrodinger equation with class of Yukawa potential for a quakonium system via series expansion method*, European J. Applied Phys. 2 (2020) 6 1-6. https://doi.org/10.24018/ejphysics.2020.2.6.26

[33]   C. O. Edet, U. S. Okorie, A. T. Ngiangia and A. N. Ikot, *Bound state solutions of the Schrodinger equation for the modified Kratzer potential plus screened Coulomb potential,l* Indian J. Phys. 19 (2019) 01477, 1-9. https://doi.org/10.1007/s12648-019-01477-9

[34]    C. O. Edet, K. O. Okorie, H. Louis, and N. A. Nzeata-Ibe, *Any l-state solutions of the Schrodinger equation interacting with Hellmann–Kratzer potential model,* Indian J. Phys. 19 (2019) 01467, 1-9. https://doi.org/10.1007/s12648-019-01467-x

[35]   E. S. William, E. P. Inyang, E. A. Thompson, *Arbitrary $\ell$-solutions of the Schrödinger equation interacting with Hulthén-Hellmann potential model*, Rev Mex Fis, 66 (2020) 6, 1-11. https://doi.org/10.31349/RevMexFis.66.730

[36]   E. P. Inyang *et al. Any State Solutions Of The Schrödinger Equation Interacting With Class Of Yukawa - Eckart Potentials*. Intl J. Innov Sci, Engr. Tech.7 (2020) 11

[37]   C. O. Edet, P. O. Amadi, U. S. Okorie, A. Tas, A. N. Ikot and G. Rampho. Solutions of Schrodinger equationand thermal properties of generalized trigonometric Poschl-Teller potential, Rev. Mex. Fis. 66 (2020) 6, 824-839. Doi: https://doi.org/10.31349/RevMexFis.66.824

[38]   A. F. Nikiforov, V. B. Uvarov. *Special Functions of Mathematical Physics*. Birkhäuser, Basel (1988)

[39]   E. P. Inyang  E. P. Inyang  J. E. Ntibi, E. E. Ibekwe, and E. S. William, *Approximate solutions of D-dimensional Klein-Gordon equation with Yukawa potential via Nikiforov-Uvarov method*, Indian J. Phys. (2020) 00097R2, 1-14. https://doi.org/10.13140/RG.2.2.32473.34406

[40]   C. A. Onate,  J. O. Ojonubah. *Relativistic and nonrelativistic solutions of the generalized Poschl-Teller  and hyperbolical potentials with some thermodynamic properties*, Intl J. Mod Phys E  24 (2015)  3 1550020, 1-16. https://doi.org/10.1142/S0218301315500202

[41]   O. Ebomwonyi, C. A. Onate O. E. Odeyemi, *Application of Formula Method for Bound State Problems in Schrödinger Equation*. J. Appl. Sci.  Environ Manage. 23(2019) 2, 323-331. https://doi.org/10.4314/jasem.v23i2.19.

[42]   W.C. Qiang, Y. Gao, R.S. Zhou, *Arbitrary l-state approximate solutions of the Hulthén potential through the exact quantization rule*, Cent. Eur. J. Phys. 6 (2008) 2, 356-362. https://doi.org/10.2478/s11534-008-0041-1

[43]   O. Bayrak, G. Kocak, I. Boztosun,  *Any l-state solutions of the Hulthén potential by the asymptotic iteration method*, J. Phys. A: Math Gen. 39 (2006) 11521-11529. https://doi.org/10.1088/0305-4470/39/37/012

[44]   S.M. Ikhdair, *An improved approximation scheme for the centrifugal term and the Hulthén potential*, Eur. Phys. J. A 39 (2009) 3, 307-314.